\documentclass[prl,floatfix,twocolumn,showpacs,superscriptaddress]{revtex4}
\usepackage{epsfig}
\usepackage{psfig}
\usepackage{epic}
\usepackage{amssymb}
\usepackage{fancyhdr}
\usepackage{fancybox}
\usepackage{shapepar}
\def\bea{\begin{eqnarray}}
\def\eea{\end{eqnarray}}
\def\be{\begin{equation}}
\def\ee{\end{equation}}

\begin{document}
\author{D.~Poilblanc}
\affiliation{Laboratoire de Physique Th\'eorique CNRS-UMR5152,
Universit\'e Paul Sabatier, F-31062 Toulouse, France }

\author{E.~Orignac}
\affiliation{Laboratoire de Physique Th\'eorique de l'\'Ecole
Normale Sup\'erieure CNRS-UMR8549, F-75231 Paris Cedex 05, France}

\author{S.R.~White}
\affiliation{Department of Physics, University of California, Irvine
CA 92697}

\author{S.~Capponi} \affiliation{Laboratoire de Physique Th\'eorique CNRS-UMR5152,
Universit\'e Paul Sabatier,
  F-31062 Toulouse, France }

\date{\today}
\title{Resonant magnetic mode in superconducting 2-leg ladders}
\begin{abstract}
The spin dynamics of a doped 2-leg spin ladder is investigated by
numerical techniques. We show that a hole pair-magnon boundstate
evolves at finite hole doping into a sharp magnetic excitation
below the two-particle continuum. This is supported by a field
theory argument based on a SO(6)-symmetric ladder. Similarities
and differences with the resonant mode of the high-T$_c$ cuprates
are discussed.
\end{abstract}

\maketitle

Spin ladders materials are built from weakly coupled ladder units
consisting of chain sub-units of spins S=1/2 (so-called ``legs'')
connected via some ``rung'' couplings and exhibit numerous
fascinating and intriguing properties~\cite{review}. Among them,
the role of the parity of the number of legs is remarkable: only
ladders with an even number of legs exhibit a spin gap (ie a
finite energy scale for triplet excitations) as seen from
different characteristic behaviors of the spin susceptibility
observed in SrCu$_2$O$_3$ and SrCu$_3$O$_5$ \cite{ladders:suscep},
typical 2-leg and 3-leg compounds respectively.

Besides exotic experimental properties, spin ladders are of
special interest for theorists as well, especially the simple
2-leg ladder: indeed, it is believed that its ground state (GS) is
a close realization of the Resonating Valence Bond (RVB) state
proposed by Anderson~\cite{RVB} in the context of high-T$_C$
superconductivity. The striking difference between ``even'' and
``odd'' ladders mentioned above can naturally be explained in this
simple picture since pairing nearest neighbor (NN) spins into spin
singlets is more easily realized on rungs with an even number of
sites. While the generic two dimensional (2D) Mott insulator (for
one electron per site) is antiferromagnetic (AF) and may evolve
into an RVB state only at finite doping, the 2-leg spin ladder
exhibits a finite spin correlation length. It is therefore an
ideal system to investigate doping into the RVB-like spin
liquid~\cite{ladder:supercond} leading to d$_{x^2-y^2}$-like
pairing~\cite{pairing}. As a matter of fact, pressure was shown to
induced superconductivity in the intrinsically doped
Sr$_2$Ca$_{12}$Cu$_{24}$O$_{41+\delta}$ ladder
material~\cite{ladder:supercond2} hence bridging the apparent gap
between 2-leg ladder and layer-based cuprates.

\begin{figure}
  \centerline{\includegraphics*[width=0.8\linewidth,angle=0]{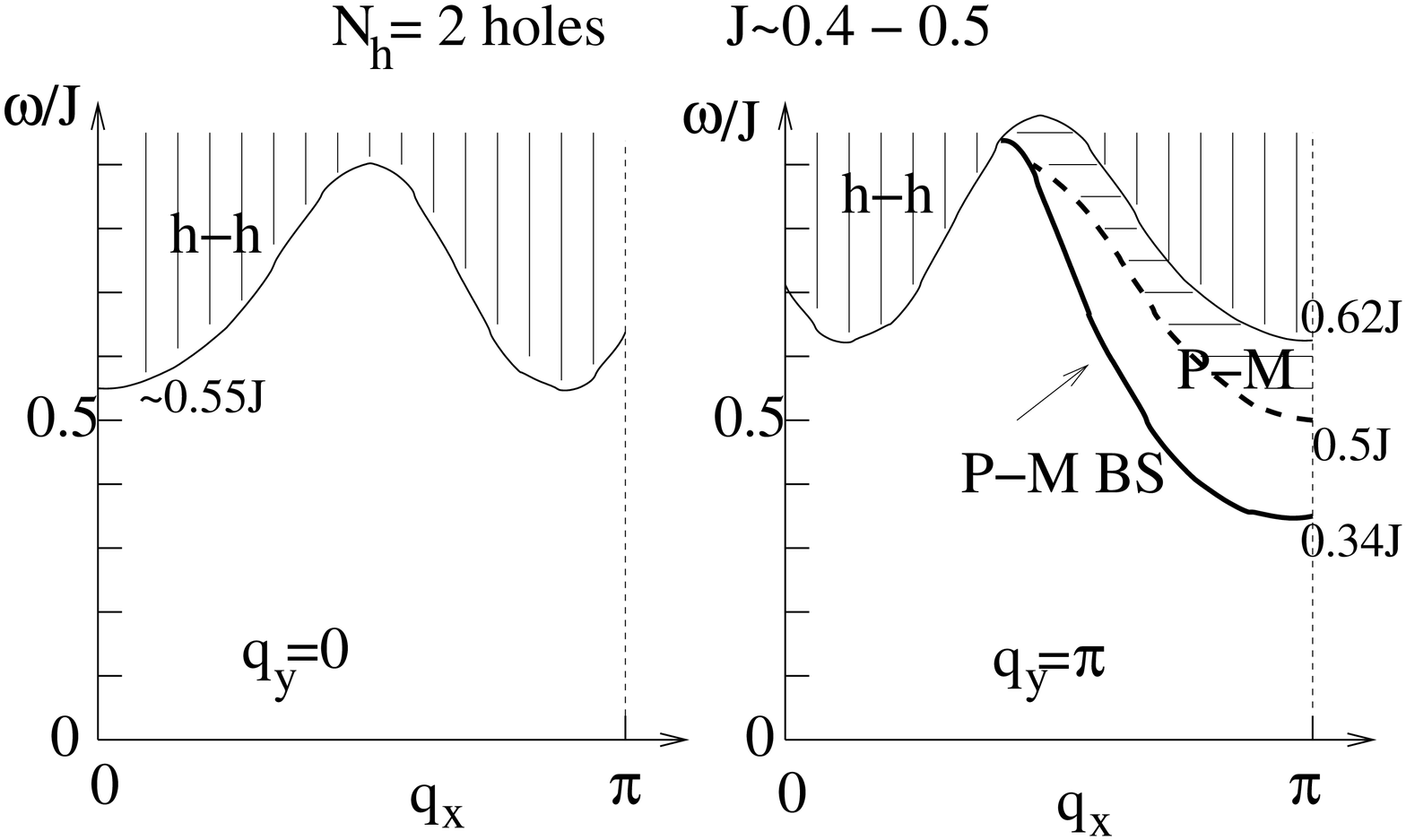}}
  \caption{\label{fig:2holes}
Schematic picture of the low-energy triplet spectrum of a spin
ladder doped with 2 holes. The left and right panels correspond to
the two transverse momenta (as written on the plots). The two-hole
continuum (h-h) and the pair-magnon (P-M) scattering continuum are
shown as hashed regions. The thick line corresponds to the
pair-magnon boundstate (P-M BS). Note that the single magnon
branch is not shown (hidden by the P-M states).}
\end{figure}

In this Letter, we investigate the spin dynamics of a 2-leg spin
ladder doped with mobile holes. In 2D superconducting cuprates, a
resonant mode at an energy around 40~meV was observed by Inelastic
Neutron Scattering (INS)~\cite{resonant,sidis}. This mode was
shown to be quite
sharp
both in energy and in momentum space (centered around the AF
wavevector) and its observation seems to be directly linked to the
appearance of superconductivity. Anomalous spectral lineshape
in photoemission experiments was also interpreted as the effect of
a coupling of the quasiparticles to a collective
mode~\cite{PES} related to the pairing interaction.
On the other
hand, in ladders, the spin dynamics remains largely to be
explored. We believe that a closer inspection, both experimentally
and theoretically, of the low energy spin excitations in such a
basic system will provide insights into the mechanism of pairing
mediated by spin fluctuations in the 2D high-T$_C$ analogs. In
this study, combining different techniques, we provide evidences
of a new magnetic mode in the two ladder system and discuss
similarities and differences with the resonant mode of the 2D
cuprates.

We shall consider here a generic 2-leg t-J ladder,
\begin{eqnarray}
\label{ham} &{\cal H}&=J_{\rm leg}\sum_{i,a} {\bf S}_{i,a} \cdot
{\bf S}_{i+1,a}
+J_{\rm rung} \sum_{i} {\bf S}_{i,1} \cdot {\bf S}_{i,2} \\
&+& t_{\rm leg} \sum_{i,a} c_{i,a}^\dagger c_{i+1,a} +t_{\rm rung}
\sum_{i,a} c_{i,a}^\dagger c_{i,a+1}+ H.C.\, , \nonumber
\end{eqnarray}
\noindent where $c_{i,a}$ are projected hole operators and $a$
(=$1,2$) labels the two legs of the ladder. Isotropic couplings,
$t_{\rm leg}=t_{\rm rung}=t$ and $J_{\rm leg}=J_{\rm rung}=J$,
will be of interest here. A value like $J=0.4$ ($t$ is set to $1$)
and dopings between 0.1 and 0.2 are typical of the above-mentioned
superconducting ladder materials. In this regime, the ladder
belongs to the Luther-Emery class with a single zero-energy charge
mode~\cite{LE,2holes1} and the dominant superconducting
correlations exhibit power-law decay~\cite{HPNSH95}.

Prior to the investigation of the spin dynamics at finite doping
it is very instructive to first consider a simpler case consisting
of two holes (forming a hole pair) immersed into an infinite (or
very large) ladder. Following previous
investigations~\cite{2holes1} and using finite size
extrapolations~\cite{2holes2}, we have tentatively sketched the
triplet excitation spectrum in Fig.~\ref{fig:2holes}. Apart from
the two expected types of excitations associated to (i) the
dissociation of a hole pair into two separate holes leading to a
two-hole continuum and to (ii) magnon-hole pair scattering states,
a boundstate of the hole pair with a magnon state
(see~\cite{2holes2} for details about the boundstate wavefunction)
can also be clearly identified. How this remarkable feature
evolves at {\it finite} doping is the crucial issue we would like
to address hereafter.

\begin{figure}
 \centerline{\includegraphics*[width=0.7\linewidth,angle=-90]{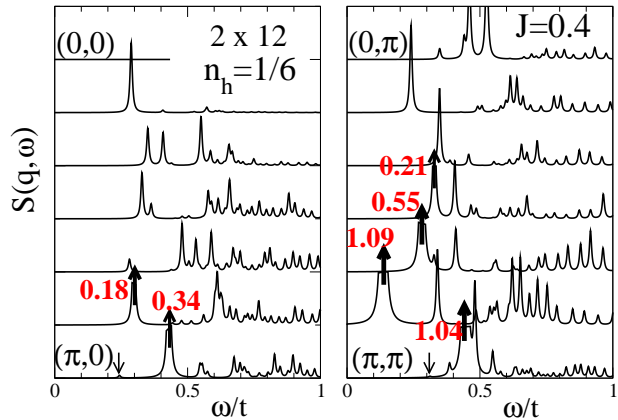}}
  \caption{\label{fig:Sqw}
Dynamic spin structure factor computed on a $2\times 12$ ladder at
$1/6$ doping and $J=0.4$. Data for $0$ and $\pi$ transverse
momenta are shown on the left and right panels respectively, and
$q_x$ varies from $0$ (top) to $\pi$ (bottom). The weights of the
largest peaks (truncated and symbolized as thick arrows) are
indicated on the plot in units where the "local"
$\omega$-integrated weight is $\frac{1}{N} \sum_{\bf q} S({\bf q})
= 1-n_h \simeq 0.833$. Down arrows indicate (almost) silent
excitations.}
\end{figure}

For that purpose, the dynamic spin structure factor
\begin{equation}
  S({\bf q},\omega)=-\frac{1}{\pi}
  \mbox{Im}[\langle\Psi_0|S_{\bf -q}^\alpha
  \frac{1}{\omega+E_0+i\eta-H}S_{\bf q}^\alpha|\Psi_0\rangle],
  \label{eq:sqw}
\end{equation}
is computed by Lanczos Exact Diagonalisations (ED) of $2\times L$
periodic ladders of size $L=12$ supplemented by a
continued-fraction technique. Typical results shown in
Fig.~\ref{fig:Sqw} bear similarities with the case of 2 holes in
Fig.~\ref{fig:2holes}. First, in the vicinity of momentum
$(\pi,\pi)$ a strong low energy incommensurate peak is observed
well separated from the higher energy excitations. Its energy
rises when ${\bf q}\rightarrow (\pi,\pi)$ where it becomes damped.
Assuming single-particle excitations~\cite{2holes1} at momenta $\pm k_{F,b}$
($b=1,2$ for bonding and anti-bonding branches, $k_{F,b}\sim \pi/2$),
this feature can naturally be assigned to a $k_{F1}+k_{F2}=\pi(1-n_h)$
spin density wave (SDW) type of excitations.
For vanishing doping,  $k_{F1}+k_{F2}\rightarrow\pi$,
corresponding to the momentum of the
hole pair-magnon boundstate.
Note also that the  boundstate in Fig.~\ref{fig:2holes}
has a minimum energy
of $0.34J\simeq 0.14$ close to the energy of the incommensurate peak
in the right panel of Fig.~\ref{fig:Sqw}.
Furthermore, the lowest triplet excitations at $q_y=0$ might be
interpreted as the sum of two single-particle excitations
with momenta around $\pm (k_{F,1},0)$ or around $\pm (k_{F,2},\pi)$
giving an excitation with total longitudinal momentum 0 or
$2k_{F,1}\simeq 2k_{F,2}$. In that case, the peaks in the left panel of
Fig.~\ref{fig:Sqw}
would signal the edge of this 2-particle continuum (or a new boundstate
with small binding energy).
Note that the lowest excitation energy at $q_y=0$
is indeed close to that of Fig.~\ref{fig:2holes}.

\begin{figure}
  \centerline{\includegraphics*[width=0.8\linewidth,angle=0]{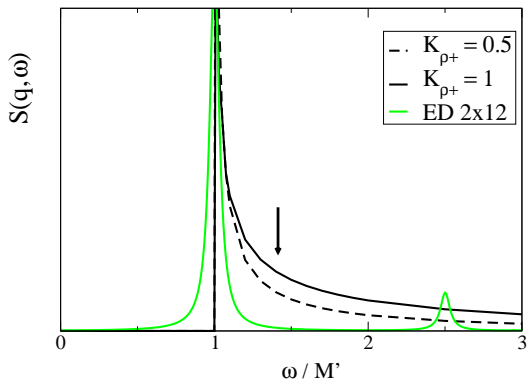}}
  \caption{\label{fig:Sqw_boson}
Contribution of the 2-kink (or 2-antikink) boundstates to the
dynamic spin structure factor in the SO(6)-symmetric ladder at
momentum ${\bf q}=(2k_F,\pi)$ (shown for 2 values of $K_{\rho+}$).
The frequency is normalized to the boundstate energy
$M'=\sqrt{2}M$. The arrow indicates the onset of a small kink-kink
continuum starting at $\omega=2M$ (not shown). Note that the ED data of
the t-J model shown on
the same scale (setting $M'$ as the finite size spin gap) exhibit
only a small number of poles ($\delta$-functions).}
\end{figure}

From the similarities between
Figs.~\ref{fig:2holes} and \ref{fig:Sqw}
it is tempting to interpret the $q_y=\pi$ feature as a collective mode
emerging below the continuum. However, the ED data do not provide
enough energy (and momentum) resolution
to investigate the shape of the spectral weight in
more details.
For this purpose, the bosonized ladder model with SO(6)$\sim$SU(4)
symmetry~\cite{Schulz,SO6} is of special interest.
Schulz~\cite{Schulz} first pointed out the
existence in this model of kink boundstates (Majorana
fermions) related to $(2k_F,\pi)$ spin density waves. Calling $M$ the mass of
elementary kinks~\cite{note_kink}, the mass of these boundstates is
$M'=M\sqrt{2}$ and the continuum starts at $2M$.
We believe that the  boundstate is, in fact, robust and
more generic and that the above feature seen in the numerics of
the t-J ladder (with lower symmetry than SO(6)) is a signature of
it. This conjecture is partially supported by the fact that
the renormalization
group flow of the bosonized ladder model is attracted to the direction of SO(6)
symmetry~\cite{Schulz}.  Let us briefly summarize the derivation of the low energy spin
dynamics within the SO(6) field-theoretic scheme. In bosonisation, the
SDW operator at $q_x\sim 2k_F$ ($=k_{F,1}+k_{F,2}$) and $q_y=\pi$
is written as $O_{SDW}=e^{i
\phi_{\rho+}(x)} N(x)$. The correlation function $\big<N(x,\tau)
N(x)\big>$ can be decomposed into\cite{konik,SU4}:
\begin{eqnarray}
&&\int d\theta |\big<0|N|B(\theta)\big>|^2
exp ( M'(i\frac{x}{v}\sinh{\theta} -\tau \cosh{\theta}))\nonumber \\
&+&\int d\theta_1 d\theta_2 |\big< 0|N|A(\theta_1)A(\theta_2)
\big>|^2 \nonumber \\ && exp( i\frac{Mx}{v} (\sinh{\theta_1} +
\sinh{\theta_2}) -M \tau (\cosh{\theta_1} +
\cosh{\theta_2}))\nonumber \\ &+& ...
\end{eqnarray}
where $|B(\theta)\big>$ is a state which contains a single
Majorana fermion of rapidity $\theta$, $|A(\theta_1)
A(\theta_2)\big>$ a state with 2 kinks of rapidity $\theta_1
\theta_2$. Since Lorentz invariance~\cite{konik} makes
$\big<0|N|B(\theta)\big>$ independent of $\theta$, the
contribution of the Majorana fermion to the correlation
function~\cite{note2} becomes $\big<N(x,\tau) N(x)\big>_{\rm
1fermion} = K_0(M' r/v)$ with $r=\sqrt{x^2 +(v\tau)^2}$. In real
space, the correlation function $\big<O_{SDW}(x,\tau)
O_{SDW}(0,0)\big>$ is  given by $(1/r)^{K_{\rho+}/2} K_0(M' r/v)$
where $K_{\rho+}$ is a non-universal constant whose inverse is the
exponent of the power law decay of the superconducting
correlations. The Fourier transform of this correlator involves a
Weber Schaefetlin integral\cite{SU4} and, eventually, one gets for
the dynamic susceptibility: $\chi_{SDW}(q,\omega) \sim
F_1(1-K_{\rho+}/4,1-K_{\rho+}/4;1; (\omega^2 -(vq)^2)/(M')^2) +
...$, where $F_1$ is the usual hypergeometric function and
$q=q_x-2k_F$. At the onset, the imaginary part of
$\chi_{SDW}(q,\omega)$, i.e. $S({\bf q},\omega)$ for ${\bf
q}=(2k_F+q,\pi)$, exhibits a power-law singularity instead of a
$\delta$-function as seen in Fig.~\ref{fig:Sqw_boson}. In the
vicinity of the onset one gets, $\chi_{SDW}(q,\omega) \sim
(1/(\omega^2-(vq)^2-M'^2))^{1-K_{\rho+}/2}$. Physically, it means
that, in e.g. an inelastic neutron scattering (INS) experiment,
holons can be excited together with the triplet boundstate, the
initial neutron momentum being distributed among all the
constituants. A complete characterization of such a singularity is
beyond the ability of present day numerical techniques but the
results of Fig.~\ref{fig:Sqw}, shown also in
Fig.~\ref{fig:Sqw_boson}, revealing a low energy pole with a large
fraction of the spectral weight accompanied by a few tiny peaks
are consistent with such a scenario.

\begin{figure}
  \centerline{\includegraphics*[width=0.8\linewidth,angle=0]{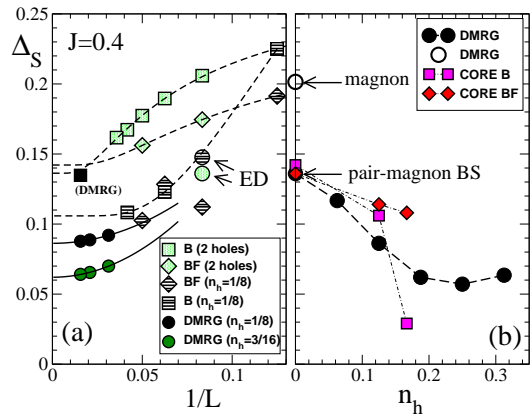}}
  \caption{\label{fig:Spin_gap}
Spin gap of the 2-leg ladder at $J=0.4$; (a) Finite size scaling
analysis for a fixed hole number (of 2) or a fixed  hole density
computed using bosonic or boson-fermion range-2 CORE hamiltonians
(see text) or DMRG methods. Dashed line are tentative fits of the
data with algebraic finite size corrections for DMRG ($\sim
1/L^2$) and exponential for CORE. (b) Variation of the spin gap as
a function of doping.}
\end{figure}

To complete our study of the spin dynamics, we have combined
complementary Density Matrix Renormalisation group (DMRG) and
Contractor Renormalisation (CORE) methods~\cite{CORE} supplemented
by finite size analysis to extract the collective mode energy. In
the CORE method the ladder is decomposed into plaquette units and
an effective interaction between adjacent plaquettes is
constructed. In the bosonic version (B), on each plaquette, only
the lowest energy undoped singlet (vacuum), triplet state (mapped
as a triplet boson) and hole pair GS (mapped as a pair boson) are
retained while in the more involved boson-fermion (BF) version the
lowest one-hole doped plaquette states are also included.
Typically, the effective B-hamiltonian is simply written as a sum
of a simple bilinear kinetic term for the bosons and a quartic
interaction. DMRG and CORE computations involving open and
periodic boundary conditions respectively give rise to different
types of finite size corrections as seen in
Fig.~\ref{fig:Spin_gap}(a). The resulting extrapolations of the
lowest triplet excitation are shown in Fig.~\ref{fig:Spin_gap}(b)
as the function of the hole doping. All extrapolations with a
fixed number of 2 holes agree very well with each other and, as
mentioned in Refs.~\cite{2holes1,2holes2}, the spin excitation in
the limit of vanishing doping is not continuous due to the binding
of the magnon with a hole pair. The main result shown by
Fig.~\ref{fig:Spin_gap}(b) is that the energy of the magnon-hole
pair state evolves continuously, at finite doping, into the
magnetic mode energy previously studied. Note that the CORE
approaches (both B \& BF versions) remain very accurate for hole
doping below 10\% but the bosonic version fails for larger doping.
Physically, this signals that the pair size becomes too large for
the spin-singlet and spin-triplet pair wavefunctions to be
correctly captured by the small subset of states kept here (unless
longer range effective interactions are included).

\begin{figure}
  \centerline{\includegraphics*[width=0.8\linewidth,angle=0]{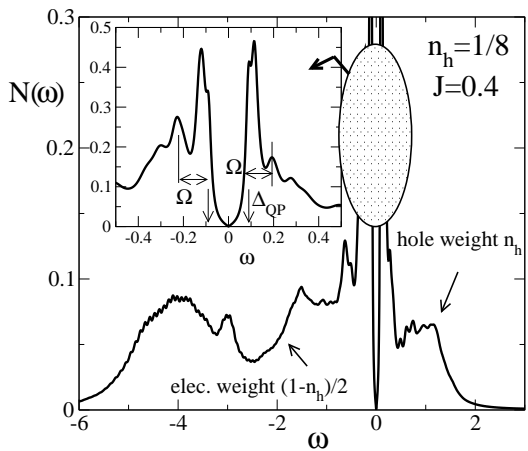}}
  \caption{\label{fig:density}
Tunnelling density of state of the 2-leg ladder computed on a
$2\times 12$ cluster. The integrated weights (sum rules) for
electrons and holes are shown on the plot. Inset: close up of the
low energy region.}
\end{figure}

Apart from singularities in $S({\bf q},\omega)$, the spin-triplet
mode is expected to have effects on the spectral functions~\cite{PES}. We
have re-examined the tunnelling density of states first reported
in Ref.~\cite{2holes1,note1} looking closely at the low energy
region. As expected, the data shown in Fig.~\ref{fig:density}
fulfill sum rules and show a quasiparticle gap $\Delta_{QP}$ which
should correspond to the kink energy $M$ of the SO(6) model. After
enlarging the low energy region one can see two sub-peaks at an
energy $\Omega\sim 0.1$ from the gap edge. This could signal
scattering of the hole or electron quasiparticle by a collective
mode. Note that the characteristic energy $\Omega$ is indeed close
to the lowest spin-triplet energy reported before (see
Fig.~\ref{fig:Sqw}).

It is interesting to compare our results to the behavior of the
resonant spin excitation observed in the 2D copper oxide
superconductors by Inelastic Neutron Scattering~\cite{resonant,sidis}.
Remarkably, this mode was observed both in under-, optimally and
over-doped materials and only below the superconducting transition
temperature, T$_C$ and approximately scales with T$_C$. While the
excitation in the 2D cuprates occurs with wavevector $(\pi,\pi)$,
an incommensurate mode is found in the ladder. Although in 2D
the mode could be interpreted as a soft mode signaling
the proximity of a magnetic instability~\cite{sachdev}, its nature
should be different in the ladder which
originates from a different class of ``parent compound'', namely
a spin liquid rather than a 2D ordered AF. Note however that, in 2D,
lower energy {\it incommensurate} spin fluctuations have also been
seen experimentally pointing towards the existence of a dispersive spin
1-collective mode~\cite{sidis}, a feature that might well be related to
our finding in the ladder.

To conclude, by investigating the spin dynamics of a doped 2-leg
spin ladder, we have shown that a hole pair-magnon boundstate
evolves into a sharp magnetic excitation at finite hole doping.
Such a feature is reproduced by a SO(6) field theory which gives a
power-law singularity at the mode energy. Analogies with the
resonant mode of the high-T$_c$ cuprates have been discussed.


DP \& SC thank  IDRIS (Orsay, France) for allocation of CPU-time
on the NEC-SX5 supercomputer. DP also thanks ETH-Z\"urich for
hospitality and T.M.~Rice for numerous discussions and comments.
SRW thanks the NSF for support under grant DMR-0311843. SC also
thanks A.~L\"auchli for help in improving some numerical codes.


\begin{thebibliography}{}
\bibitem{note_kink} These massive excitations (of mass $M$) carry
a spin 1/2 and are kinks of the fields characterizing (i) the
total spin density (spinon), (ii) the difference of  spin density
and (iii) of charge current between the bonding ($q_y=0$) and
antibonding ($q_y=\pi$) bands and form a $4$ dimensional
representation of SU(4). Their boundstates, the Majorana
fermions, carry either spin-0 or spin-1 and form a 6 dimensional
representation of SU(4).
\bibitem{note2} The two-kink contribution to the correlation function
for $\omega > 2M$ can also be computed; see E.~Orignac (in
preparation).
\end{thebibliography}
\end{document}